# Basic Chemical Models of Nonideal Atomic Plasma

A. L. Khomkin[*], I. A. Mulenko[**], and A. S. Shumikhin[*]

[*]Joint Institute of High Temperatures, Russian Academy of Sciences,
Moscow, 125412 Russia

[**]Ukrainian State Marine Technical University, Nikolaev, Ukraine

The concept of "basic chemical models" is introduced, which is new from the standpoint of the physics of nonideal atomic plasma. This concept is based on the requirement of full conformity of the expression for free energy in the chemical model of plasma to exact asymptotic expansions obtained in the grand canonical ensemble within the physical model of plasma. The thermodynamic functions and equations of state and ionization equilibrium are obtained for three basic chemical models differing from one another by the choice of the atomic partition function. Comparison is made with the experimental results for nonideal plasma of cesium and inert gases. It is demonstrated that the best fit to experiment is shown by the results obtained using a basic chemical model with atomic partition function in the nearest neighbor approximation with classical determination of the size of excited atom.

## INTRODUCTION

We will treat an atomic hydrogen-like plasma consisting of $N_e$ electrons, $N_i$ ions, and $N_a$ atoms and located in a volume $V$ at a temperature $T$. The thermodynamic properties of such plasma may be described within both "chemical" (free electrons, ions, and atoms) [1] and "physical" (electrons and nuclei) [2-4] models of plasma. Assuming that the gas of atoms is ideal and that the free charges weakly interact with one another the free energy $F$ of the system being treated will have the following form within the "chemical model":

$$\beta F = -N_e \ln(\frac{2eV}{N_e \lambda_e^3}) - N_i \ln(\frac{eV}{N_i \lambda_i^3}) - N_a \ln(\frac{e\Sigma_a V}{N_a \lambda_a^3}) - (N_e + N_i)\Delta f, \qquad (1)$$

where $\beta = 1/kT$ is the inverse temperature; $\lambda_l = \left(2\pi\hbar^2\beta/m_l\right)^{1/2}$ is the thermal wavelength of a particle of the sort $l = e, i, a$; $\Sigma_a$, - is the internal atomic partition function; $e$ - is the base of natural logarithm; and $\Delta f$ - is the correction to the free energy of an ideal-gas mixture in temperature units per particle, caused by the interaction of free electrons and ions with one another. Equation (1) describes a system of charges in a fairly wide range of pressures and temperatures from an ideal gas of atoms to fully ionized weakly nonideal plasma.

The arbitrary choice of the quantities $\Sigma_a$ and $\Delta f$ leads to numerous versions of "chemical model" of plasma (1), while the "physical" model [2-4] is free of this indeterminacy. It follows from the literature that the problem of consistent description of the thermodynamic properties of nonideal plasma using physical and chemical models has existed in the physics of nonideal plasma until very recently [5].

Without placing restrictions on the generality of further computations, we will assume that the dimensionless correction $\Delta f$ is a function of only the plasma parameter $\Gamma = \beta e^2 \sqrt{4\pi\beta e^2 (n_e + n_i)}$ ($n_l$ denotes the concentrations of particles of the sort $l$), and the atomic partition function depends symmetrically on the concentrations of free charges $\partial\Sigma_a/\partial n_e = \partial\Sigma_a/\partial n_i = \partial\Sigma_a/\partial n_{ei}$ [1].

An analytical expression relating the correction $\Delta f$ to the atomic partition function $\Sigma_a$ was derived in [1] within these assumptions,

$$\Delta f = \frac{\Gamma}{3} - \frac{n_e n_i}{n_e + n_i}\lambda_e^3 \frac{1}{2}\left(\Sigma_a - \Sigma_{PL}\right) \qquad (2)$$

where

$$\Sigma_{PL} = \sum_{k=1}^{\infty} 2k^2 \left(\exp\left(\beta Ry/k^2\right) - 1 - \beta Ry/k^2\right) \qquad (3)$$

is the Planck–Larkin partition function, and $Ry$ is the ionization potential of hydrogen atom.

Relation (2) is derived as follows. Expression (1) for free energy is used to make a transition to the grand canonical ensemble and to derive the expression for the grand thermodynamic potential $\Omega_{CM}$, which corresponds to chemical model (1). We equate this expression to exact asymptotic expansion for the grand thermodynamic potential $\Omega_{PM}$, obtained within the physical model [2-4],

$$\Omega_{CM} = \Omega_{PM} \qquad\qquad\qquad (4)$$

change in relation (4) from activities to concentrations, which in fact implies the inverse transition to the small canonical ensemble, and calculate relation (2) which is written in the final form within terms of the order of $\Gamma^2$.

We suggest that the chemical models constructed using relation (2) should be referred to as "basic chemical models". Their distinguishing feature is the exact conformity of expression (1) for free energy to the results obtained using the "physical model" for the grand thermodynamic potential [2-4]. As a result, it turned out possible to formulate, for the first time, the theoretical procedure for matching the calculation of the thermodynamic properties of weakly nonideal plasma within the physical and chemical models. It turns out that almost none of tens of chemical models employed in the literature is basic within our definition (2).

We used the Planck–Larkin approximation (PLA) and the nearest neighbor approximation (NNA) for the calculation of the atomic partition function to derive relations for all thermodynamic functions and equations of state and ionization equilibrium. Comparison was made with the experimental results of [5] for cesium, argon, and xenon plasma.

**1. Thermodynamic functions of atomic plasma.** We will use the standard thermodynamic relations

$$\beta P = -\frac{\partial \beta F}{\partial V}, \qquad\qquad\qquad ` \qquad\qquad (5)$$

$$\beta E = \frac{\partial \beta F}{\partial \beta} \qquad\qquad\qquad\qquad (6)$$

to derive, from Eq. (1),

$$\beta P = (n_e + n_i)(1 - \Delta p) + n_a, \tag{7}$$

$$\frac{\beta E}{V} = \frac{3}{2}(n_e + n_i + n_a) - n_a E_a - (n_e + n_i)\Delta e, \tag{8}$$

and use the relation $H = E + PV$ to find

$$\frac{\beta H}{V} = \frac{5}{2}(n_e + n_i + n_a) - n_a < E_a > - (n_e + n_i)\Delta h. \tag{9}$$

Here,

$$\Delta p = \Gamma \frac{1}{2}\frac{\partial \Delta f}{\partial \Gamma} + n_a \frac{\partial \ln \Sigma_a}{\partial n_{ei}}, \tag{10}$$

$$\Delta e = \Gamma \frac{3}{2}\frac{\partial \Delta f}{\partial \Gamma}, \tag{11}$$

$$E_a = \frac{\partial \ln \Sigma_a}{\partial \beta}. \tag{12}$$

$$\Delta h = 2\Gamma \frac{\partial \Delta f}{\partial \Gamma} + n_a \frac{\partial \ln \Sigma_a}{\partial n_{ei}}. \tag{13}$$

The concentrations of electrons, ions, and atoms are related by the Saha formula with the decrease in the ionization potential $\Delta I$:

$$n_a = n_e n_i \frac{\lambda_e^3}{2}\Sigma_a \exp(-\beta \Delta I), \tag{14}$$

where

$$\beta \Delta I = 2\Delta f + \Gamma \frac{\partial \Delta f}{\partial \Gamma} + 2n_a \frac{\partial \ln \Sigma_a}{\partial n_{ei}} \tag{15}$$

Resultant relations (7–15) fully define the thermodynamics and the equations of state and ionization equilibrium of nonideal atomic plasma within chemical model (1). It follows from Eqs. (10–15) that, in order to perform concrete calculations, we must determine two quantities $\Delta f$ and $\Sigma_a$. As is observed in [1], it is this double indeterminacy that eventually gives rise to tens of chemical models of nonideal atomic plasma [6].

Relation (2) markedly reduces this indeterminacy, because the quantities $\Delta f$ and $\Sigma_a$ turn out to be related to each other. In the case of basic chemical models, it is sufficient to select the atomic partition function, after which we can use Eq. (2) to obtain the correction to free energy and Eqs. (10–15) to obtain the remaining parameters.

**2. Basic chemical models of nonideal atomic plasma.** When basic chemical models are used, two quantities are of fundamental importance, which are generally not treated in the traditional theory of nonideal plasma, namely:

$$\Delta\Sigma = \Sigma_a - \Sigma_{PL} \tag{16}$$

and

$$\Delta_{\Sigma} = n_a \frac{\partial \ln \Sigma_a}{\partial n_{ei}}. \tag{17}$$

All thermodynamic corrections and the decrease in the ionization potential, which are given by Eqs. (2) and (10–15), are expressed in terms of these quantities. We will use two models for the atomic partition function in order to calculate these quantities. These models include, first of all, the Planck–Larkin approximation which is most frequently employed in the physics of nonideal plasma, because expression (3) arises during the calculation of the converging part of the second group coefficient within the physical model and, therefore, this coefficient is often assigned the meaning of the atomic partition function. The second model for the atomic partition function involves the use of the nearest neighbor approximation (NNA) to calculate effective populations. It is very popular with authors of astrophysical papers [7]; in our opinion, preference must be given to this latter model [1]. Within this model, we have, for the atomic partition function:

$$\Sigma_{NNA} = \sum_{k=1}^{\infty} 2k^2 \exp\left(\beta Ry/k^2\right) \omega_k. \tag{18}$$

Here, $\omega_k$ is the Poisson probability of the absence of charged particles in the sphere corresponding to the size of atom in the state with the main quantum number $k$,

$$\omega_k = \exp\left[-\left(n_e + n_i\right)\frac{4\pi}{3}r_k^3\right], \quad r_k = a_0 k^2 \delta. \tag{19}$$

For further computations, we will examine two options for determining the size of atom, namely, quantum and classical ones, to which the values of $\delta = 1$ and $\delta = 2$ correspond, respectively. All relations for corrections to thermodynamic functions will be derived for an arbitrary value of $\delta$.

We will construct a basic chemical model with the atomic partition function in the Planck–Larkin approximation given by Eq. (3),

$$\Sigma_a = \Sigma_{PL}. \tag{20}$$

In this case, $\Delta\Sigma = 0$ in accordance with Eq. (16), and $\Delta_\Sigma = 0$ because $\Sigma_{PL}$ does not depend on density. It follows from formula (2) that $\Delta f = \Gamma/3$, and Eqs. (10–15) give the following corrections to thermodynamic functions:

$$\Delta p = \Gamma/6 \tag{21}$$

$$\Delta e = \Gamma/2 \tag{22}$$

$$\Delta h = 2\Gamma/3 \tag{23}$$

$$\beta\Delta I = \Gamma \tag{24}$$

The treated version of base chemical model is classical in a sense. This version uses the results of the Debye theory for the charge energy in fully ionized plasma [8]. It is this particular scheme of calculation of corrections for the nonideality of free charges in a partially ionized plasma that is suggested in the majority of monographs on plasma physics; in so doing, the question of the partition function as a rule remains open. According to our approach, it is the Planck–Larkin partition function that must be employed. The use of the set of corrections (21–24) along with some other partition function is illegitimate, because the results obtained using the physical and chemical models will not agree, and the thus obtained chemical model will not be basic. We will now examine another option of basic chemical model using the NNA for the calculation of partition function.

In order to find $\Delta\Sigma$, we will use the technique described in [1] and based on identical transformation of the atomic partition function $\Sigma_a$,

$$\Sigma_a = \sum_{k=1}^{\infty} 2k^2 \left[\exp(\beta Ry/k^2) - 1 - (\beta Ry/k^2)\right]\omega_k +$$
$$+ \sum_{k=1}^{\infty} 2k^2 \left[1 + (\beta Ry/k^2)\right]\omega_k \qquad (25)$$

We will substitute $\omega_k \to 1$ in the first term of Eq. (25), and this term will then transform to the Planck–Larkin partition function; in the second term, we will change from summation to integration. As a result, we derive

$$\Delta\Sigma = \Sigma_a - \Sigma_{PL} = \int_0^{\infty} 2k^2 \left[1 + (\beta Ry/k^2)\right]\omega_k \qquad (26)$$

After the integration of Eq. (26) with $\omega_k$ of the form of (19), we derive, for an arbitrary value of δ, $\Delta\Sigma_{NNA}$:

$$\Delta\Sigma_{NNA} = \Sigma_{NNA} - \Sigma_{PL} = \frac{1}{\sqrt{3\delta^3}} \left(\frac{1}{4\pi(n_e + n_i)a_0^3}\right)^{1/2} \Gamma\left(\frac{1}{2}\right) +$$
$$+ \frac{\beta \cdot Ry}{3\sqrt{\delta}} \left(\frac{3}{4\pi(n_e + n_i)a_0^3}\right)^{1/6} \Gamma\left(\frac{1}{6}\right) \qquad (27)$$

where $\Gamma(x)$ - is the gamma function.

We substitute Eq. (27) into (2) to derive

$$\Delta f_{NNA} = \frac{\Gamma}{3}(1 - \frac{\pi\sqrt{6}}{16}\frac{1}{\delta^{3/2}}) - \frac{\sqrt{2\pi}}{96}\frac{3^{1/6}}{\delta^{1/2}}\Gamma^{5/3}\Gamma\left(\frac{1}{6}\right). \qquad (28)$$

In calculating $\Delta_\Sigma$, we take some simplifying assumptions which are associated with the fact that the final results will be given in the form of expansion with respect to the parameter $\Gamma$:

$$\Delta_\Sigma = n_e n_i \frac{\lambda_e^3}{2} \Sigma_a \exp(-\beta\Delta I) \frac{1}{\Sigma_a} \frac{\partial\Sigma_a}{\partial n_{ei}} \approx n_e n_i \frac{\lambda_e^3}{2} \frac{\partial\Delta\Sigma}{\partial n_{ei}}. \qquad (29)$$

In writing the approximate equality in Eq. (29), we have taken into account the fact that the value of $\beta \Delta I$ is proportional to the nonideality parameter $\Gamma$ and that its inclusion will lead to terms of the order of $\Gamma^2$ and higher in the final expressions for $\Delta_\Sigma$. In substituting $\Delta \Sigma$ for $\Sigma_a$, we ignore the dependence on density in the first term of the right-hand part of Eq. (25) [1] and use relation (26) for the calculation of $\Delta \Sigma$. We substitute Eq. (27) into (29) to derive

$$\Delta_\Sigma^{NNA} = -\frac{\sqrt{\pi}}{16}\left(\frac{1}{\sqrt{6}}\frac{\Gamma}{\delta^{3/2}}\Gamma\left(\frac{1}{2}\right) + \frac{3^{1/6}}{18\sqrt{2\delta}}\Gamma^{5/3}\Gamma\left(\frac{1}{6}\right)\right) \tag{30}$$

We use the obtained expressions for $\Delta f_{NNA}$ and $\Delta_\Sigma^{NNA}$ to derive from Eqs. (10–15), for a basic chemical model with the partition function in the NNA,

$$\Delta p_{NNA} = \frac{\Gamma}{6}\left(1 - \frac{\pi\sqrt{6}}{8\delta^{3/2}}\right) - \frac{\sqrt{2\pi}\cdot 3^{1/6}}{96}\frac{\Gamma^{5/3}}{\delta^{1/2}}\Gamma\left(\frac{1}{6}\right) \tag{31}$$

$$\Delta e_{NNA} = \frac{\Gamma}{2}\left(1 - \frac{\sqrt{6}\pi}{16\cdot\delta^{3/2}}\right) - \frac{5\sqrt{2\pi}\cdot 3^{1/6}}{192}\frac{\Gamma^{5/3}}{\delta^{1/2}}\Gamma\left(\frac{1}{6}\right) \tag{32}$$

$$\Delta h_{NNA} = \frac{2}{3}\Gamma\left(1 - \frac{5\sqrt{6}\pi}{64\cdot\delta^{3/2}}\right) - \frac{7\sqrt{2\pi}\cdot 3^{1/6}}{192}\frac{\Gamma^{5/3}}{\delta^{1/2}} \tag{33}$$

$$\beta\Delta I_{NNA} = \Gamma\left(1 - \frac{\pi\sqrt{6}}{12\delta^{3/2}}\right) - \Gamma^{5/3}\frac{\sqrt{2\pi}3^{1/6}}{24\delta^{1/2}}\Gamma\left(\frac{1}{6}\right) \tag{34}$$

The dependence of corrections (31–34) on the parameter $\delta$ makes it possible to trace the part played by excited states in the thermodynamics of basic chemical models. The NNA exponentially reduces the contribution made to the atomic partition function by the excited states whose volume exceeds the volume of the Wigner-Seitz cell. By reducing the parameter $\delta$, we cause an increase in the number of excited states which make a contribution to the atomic partition function. In so doing, as is seen from relations (31–34), all corrections decrease and, for some value of $\delta$, even change sign. Because the main contribution to $\Delta\Sigma$ is that by highly excited states, the value of $\delta = 2$ appears to be most reasonable;

this corresponds to the choice of atomic volume equal to the volume of the maximal classically accessible region of motion of bound electron [9]. Another important corollary of relations (31–34) is the significant difference of obtained corrections for the nonideality of free charges of atomic plasma from the results of the classical Debye theory (21–24).

Figures 1–4 give the obtained corrections and the decrease of the ionization potential for three basic chemical models treated by us as functions of the parameter $\Gamma$. Note an important fact that the basic chemical model with the atomic partition function in the NNA exhibits a much smaller contribution by the effects of free charge interaction to the thermodynamics of partially ionized plasma compared to the basic chemical model with the Planck–Larkin partition function. The physical reasons for this effect have been discussed in [1]. This result provides a theoretical explanation for the experimentally observed "ideal" behavior of nonideal plasma. The energy of Coulomb interaction of free charge in partially ionized plasma turns out to be significantly lower than the temperature (see Fig. 3), although the formally calculated value of the nonideality parameter is high. Figure 5 illustrates the behavior of the $\Delta_\Sigma$ correction which has the physical meaning of the fraction of volume taken up by atoms of opposite sign. Indeed, $\partial \omega_k / \partial n_{ei} = -v_k \omega_k$ where $v_k = 4\pi r_k^3 / 3$ is the volume of atom in the state $k$.

**3. Comparison of calculation and experimental results. Problem of consistency between the caloric and thermal equations of state for nonideal atomic plasma.** We will treat the results of shock-wave experiments performed in cesium, argon, and xenon plasma and described by Fortov and Yakubov [5]. First of all, we will treat the isochore of cesium for a specific volume $v = 200 \, cm^3 / g$. Figure 6 gives comparison of experimental data (shaded region) and calculated curves for three basic chemical models with different partition functions, namely, those in the NNA and in the PLA. One can see that the maximal deviation from experiment is exhibited by the results of calculation by the basic chemical model with the partition function in the PLA and with Debye corrections for free charge

interaction. By the way, this inference was made previously [5]. Agreement with experiment is improved significantly when the NNA with $\delta = 2$ is used. Note that it is very difficult to attain even a slight shift of the calculated curve toward the experimental region in the case of selected coordinates.

A large body of experimental data have been obtained by now in shock-wave experiments performed in plasma of cesium and inert gases, where four thermodynamic parameters could be measured, namely, enthalpy $H$, pressure $P$, volume $V$, and temperature $T$. In comparing various theories with experiment, a certain inconsistency was observed when using the thermal ($P$, $V$, $T$) and caloric (for example $P$, $V$, $H$) equations of state [5]. Reaching agreement between theory and experiment within some model when using the thermal equation of state leads to difference between theory and experiment when using the caloric equation of state derived within the same model. It was only the approximation of ideal-gas mixture of electrons, ions, and atoms that produced reasonable agreement between experiment and the results of calculations by the thermal and caloric equations of state. One can say that it was this inference that turned out to be the most unexpected result of experiments in [5] performed for plasma with significant Coulomb nonideality. The results (26–29) obtained by us for a basic chemical model with the partition function in the NNA provide a theoretical explanation for this inference. The consistent inclusion of the contribution by highly excited bound states, defined by $\Delta\Sigma$ according to Eq. (16), results in a marked decrease in the magnitude of corrections for free charge interaction (2). In addition, one must take into account the density derivatives of the atomic partition function according to Eq. (17), as was emphasized in [10]. This problem was discussed in more detail in [11]. The term $\Delta_{\Sigma}$, though small, causes a marked reduction of the effects of Coulomb nonideality (10, 13, 15).

Ten sets of $P$, $V$, $H$, $T$ experimental data for cesium and five sets each for argon and xenon were selected for comparison with experiment. These data are given in the monograph [5]. All calculations were performed by the standard scheme: two measured quantities are used to determine the third one for which experimental data are also available. The calculations were performed within two basic chemical models, namely, those using the Planck–Larkin partition function and the NNA with $\delta = 2$.

Given in Figs. 7–10 are the results of comparison of theory and experiment for cesium (Figs. 7a–10a), argon (Figs. 7b–10b), and xenon (Figs. 7c–10c) plasma in the form of dependence of the ratio of quantities calculated within some or other model on the parameter $\Gamma$. In Fig. 7, the temperature is obtained by measured values of enthalpy $H$ and pressure $P$. In Figs. 8 and 9, the volume is calculated by $H$, $P$ and $P, T$, respectively. In Fig. 10, the enthalpy is the calculated value, and the temperature and pressure are the initial values. Therefore, both the thermal and caloric equations of state are used. The use of the basic chemical model with the Planck–Larkin partition function and Debye corrections, especially in the case of argon and xenon plasma, demonstrates poor agreement with experiment. For some experimental points in the case of argon plasma (one point) and xenon plasma (four points), the calculation within this option of basic chemical model cannot be performed at all, because these points fall into the region of instability of the model. This instability is associated with the change of sign of the derivative $\partial n_a / \partial n_e$, which gives grounds for predicting the existence of phase transition in plasma. In Figs. 7–10, the vertical dotted arrows indicate the transition to the instability region. The basic chemical model in the NNA enables one to perform calculations for the entire array of experimental points. Figures 7–10 give the calculation results only for the option of NNA with $\delta = 2$, which was selected from comparison of theory with the experimentally obtained isochore of cesium plasma for $v = 200 \, cm^3 / g$. During transition from one basic chemical model to another, the plasma composition changes, because the calculation of this composition involves the use of different partition functions and different expressions for reduction of the ionization potential (24) and (34). The basic chemical model with the Planck–Larkin partition function gives higher values of the free electron concentration compared to the basic chemical model with the partition function in the NNA. This fact results in different values of the parameter of Coulomb nonideality $\Gamma$ for one and the same set of initial thermodynamic parameters. One can see in Figs. 7–10 that the calculated points corresponding to the partition function in the NNA lie to the left of the points obtained using the Planck–Larkin partition function, because these former points have lower values of the parameter $\Gamma$ corresponding to them. The basic chemical model with the partition function in the NNA is adequately consistent with experiment in the entire experimental range, especially for the case of cesium plasma. A systematic difference between theory and experiment for xenon plasma observed in some options (Figs. 8c–10c) is possibly associated with the effect of interatomic interaction.

## CONCLUSIONS

A new concept of basic chemical models of nonideal atomic plasma has been introduced. Basic chemical models provide for consistency of the theoretical results between the chemical and physical models. Two options of basic chemical model have been examined, those using the partition function in the PLA and in the NNA. A consistent use of the NNA to determine the atomic partition function makes it possible to explain the experimentally observed effect of significant overestimation of the contribution made by Coulomb interaction to the thermodynamic functions of partially ionized plasma, as given by the Debye theory developed for fully ionized plasma. Comparison has been made of the theoretical and experimental results. Within the basic chemical model with the atomic partition function calculated in the NNA, one can reach adequate agreement with the experimental data using both the thermal and caloric equations of state.

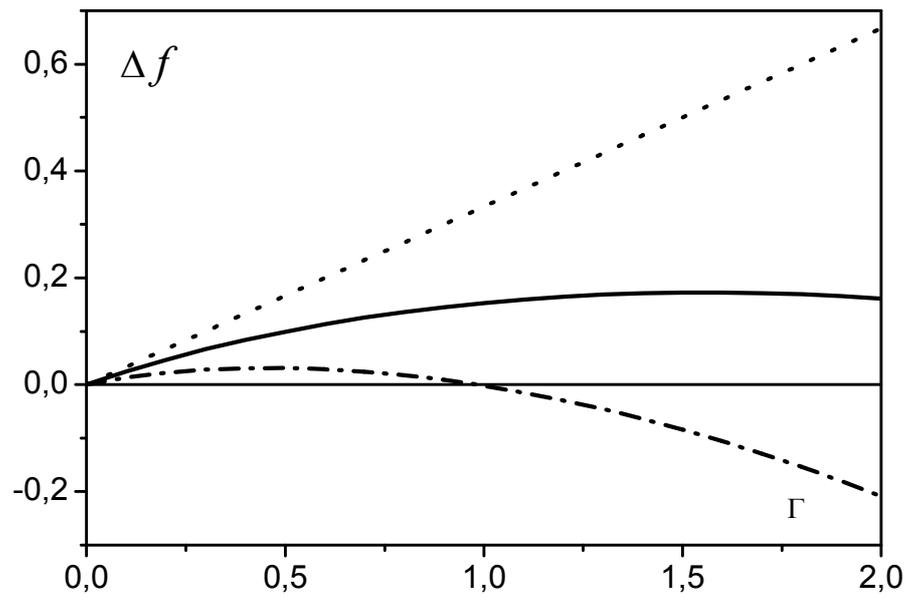

Fig. 1. Corrections to free energy $\Delta f$ as a function of the plasma nonideality parameter $\Gamma$ for basic chemical models with different atomic partition functions: dotted line, PLA; dot-and-dash line, NNA, $\delta = 1$; solid curve, NNA, $\delta = 2$.

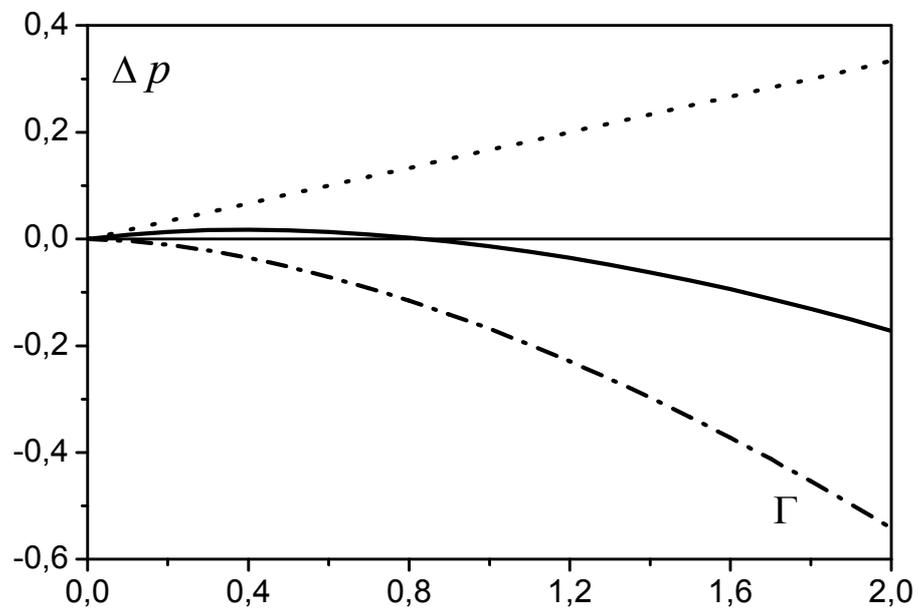

Fig. 2. Corrections to pressure $\Delta p$. Designations are as in Fig. 1.

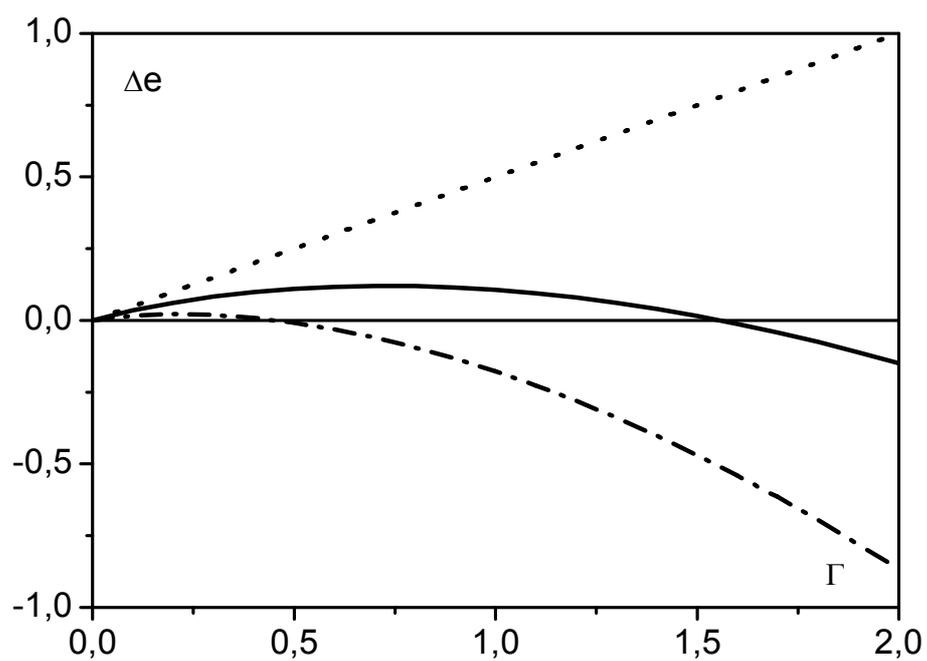

Fig. 3. Corrections to free charge energy $\Delta e$. Designations are as in Fig. 1.

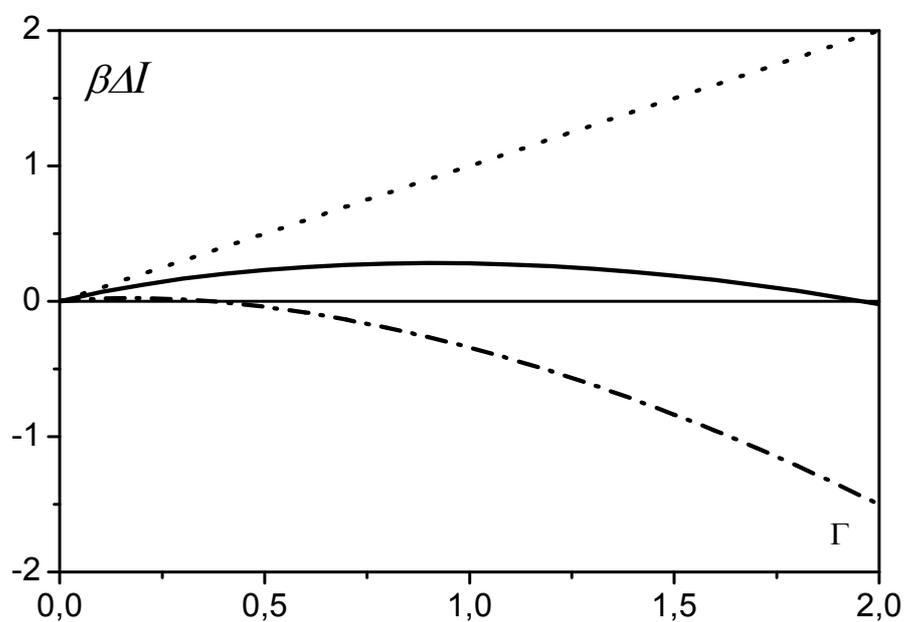

Fig. 4. Decrease in the ionization potential, related to temperature, $\beta \Delta I$. Designations are as in Fig. 1.

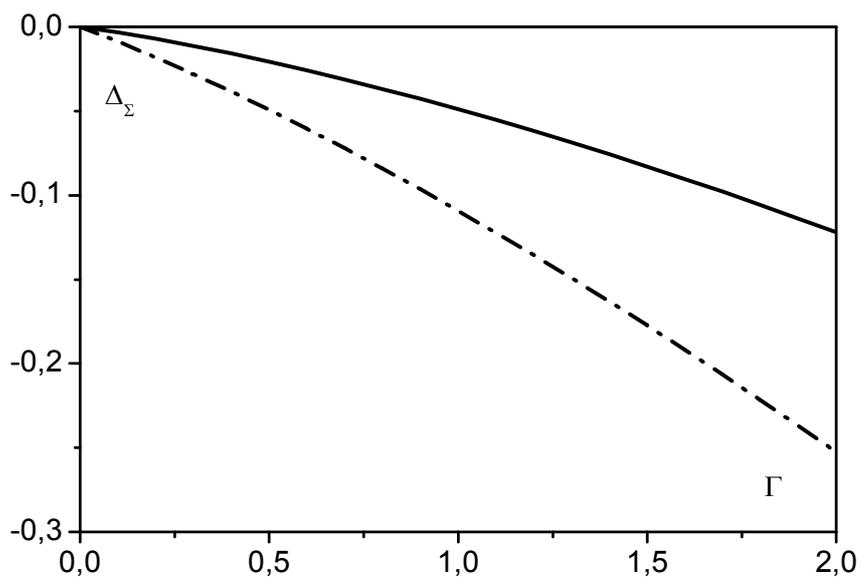

Fig. 5. The fraction of unit volume $\Delta_\Sigma$ taken up by atoms with opposite sign.

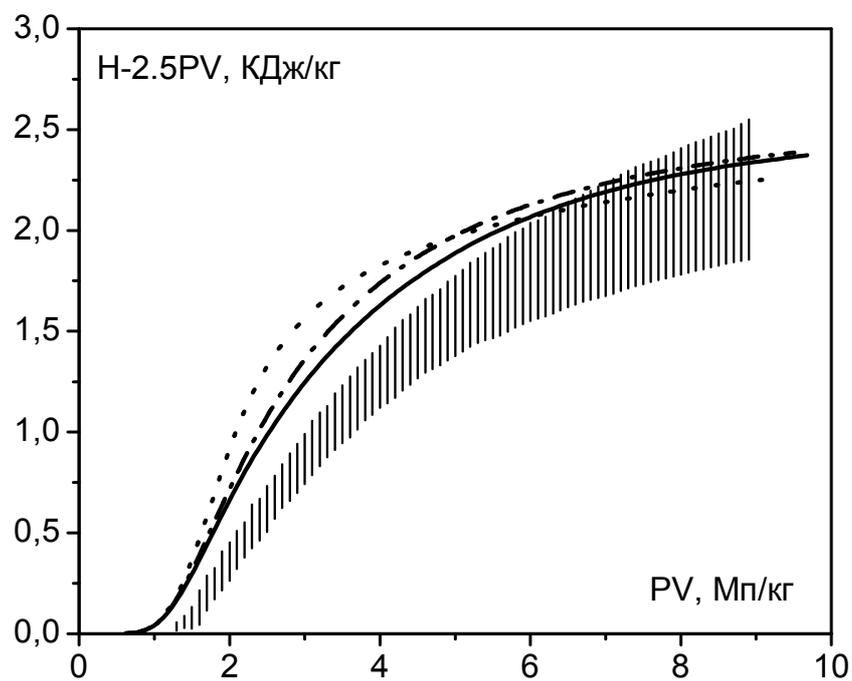

Fig. 6. Isochore of cesium plasma, V=0.2 kg/m³. Shaded region − experimental data, curves − calculation (designations are as in Fig. 1). [H-2.5PV: kJ/kg; PV: MPa/kg ]

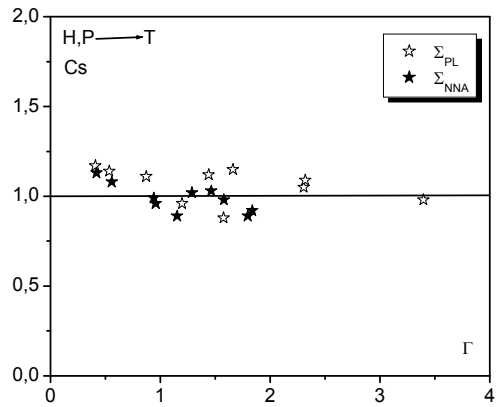
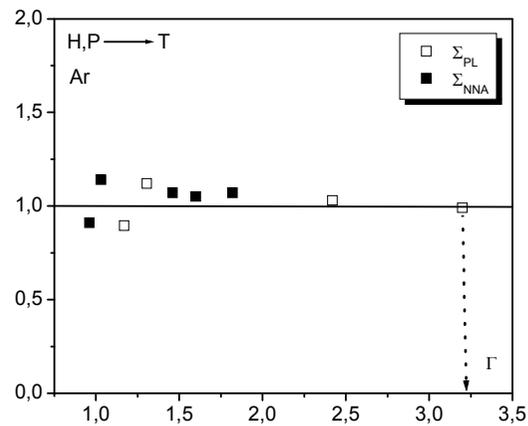
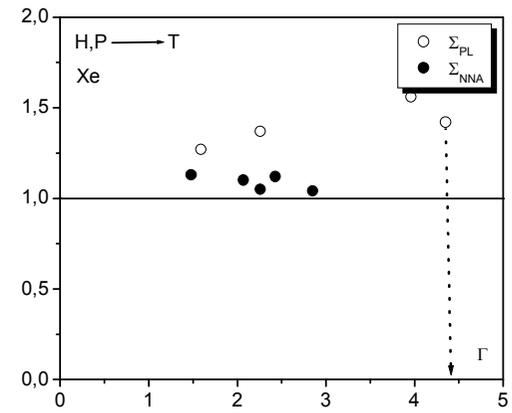

Fig. 7. The ratio of the calculated value of temperature to the experimentally obtained value (by the experimental data for $H$ and $P$): a - cesium plasma, b - argon plasma, c - xenon plasma.

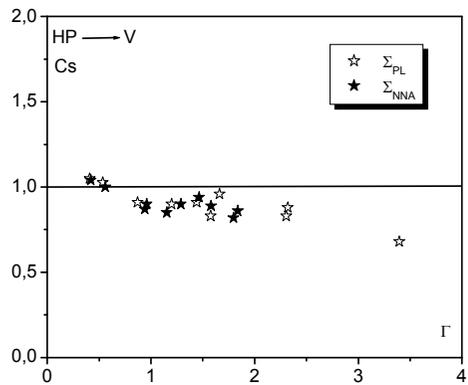
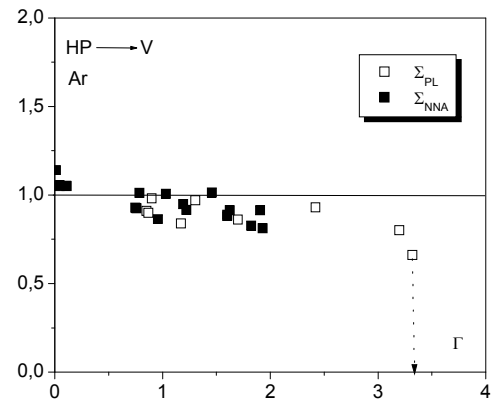
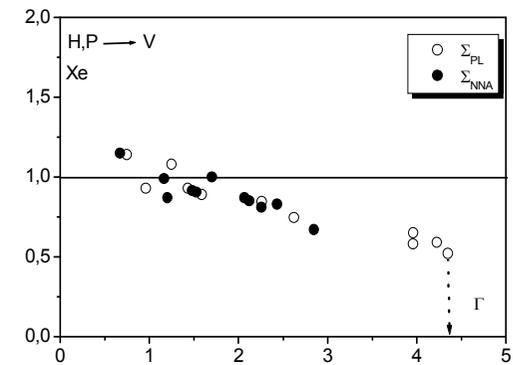

Fig. 8. The ratio of the calculated value of plasma volume to the experimentally obtained value (by the experimental data for $H$ and $P$): a - cesium plasma, b - argon plasma, c - xenon plasma.

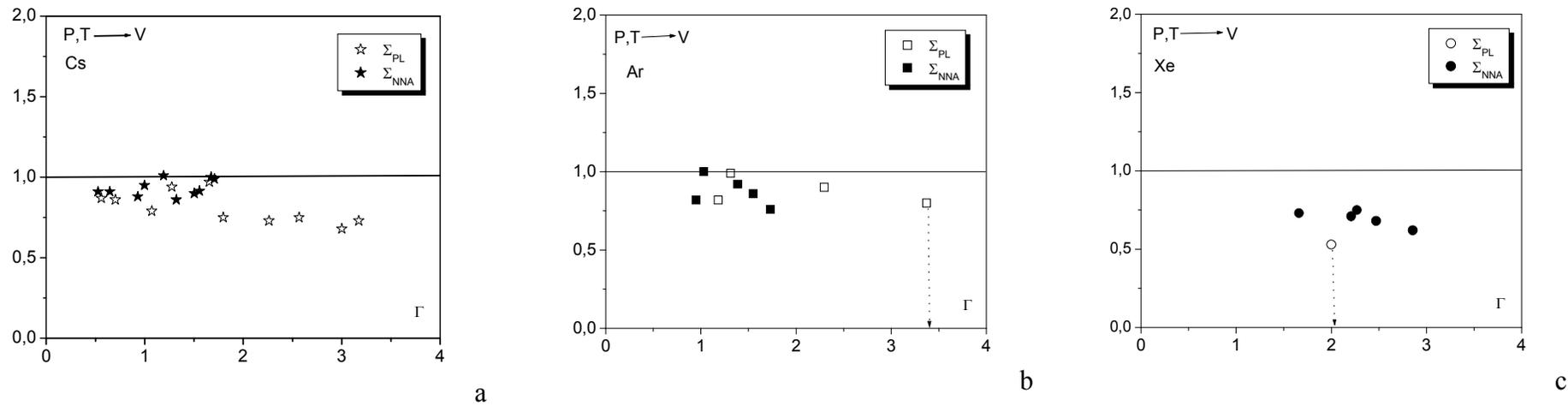

Fig. 9. The ratio of the calculated value of plasma volume to the experimentally obtained value (by the experimental data for $P$ and $T$): a - cesium plasma, b - argon plasma, c - xenon plasma.

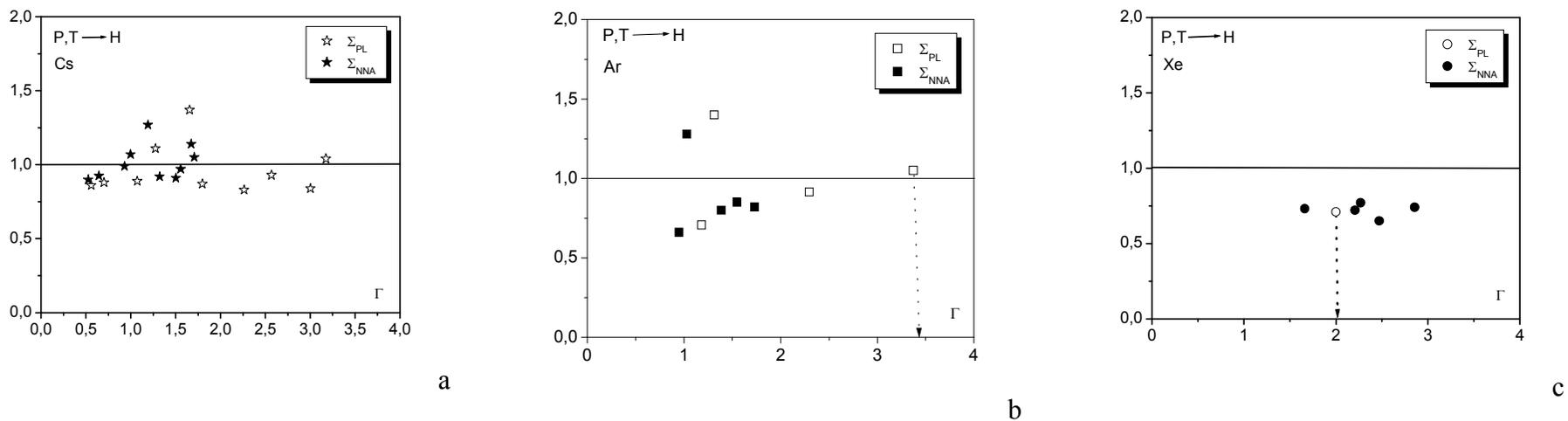

Fig. 10. The ratio of the calculated value of enthalpy to the experimentally obtained value (by the experimental data for $P$ and $T$): a - cesium plasma, b - argon plasma, c - xenon plasma.